\documentclass[12pt]{article}
\usepackage{amsmath,amsfonts,amssymb}

\begin{document}
\thispagestyle{empty}

\def\theequation{\arabic{section}.\arabic{equation}}
\def\a{\alpha}
\def\b{\beta}
\def\g{\gamma}
\def\d{\delta}
\def\dd{\rm d}
\def\e{\epsilon}
\def\ve{\varepsilon}
\def\z{\zeta}
\def\B{\mbox{\bf B}}\def\cp{\mathbb {CP}^3}

\newcommand{\h}{\hspace{0.5cm}}

\begin{titlepage}
\vspace*{1.cm}
\renewcommand{\thefootnote}{\fnsymbol{footnote}}
\begin{center}
{\Large \bf String solutions in $AdS_3\times S^3\times T^4$ with
NS-NS B-field}
\end{center}
\vskip 1.2cm \centerline{\bf Changrim  Ahn and Plamen Bozhilov
\footnote{On leave from Institute for Nuclear Research and Nuclear
Energy, Bulgarian Academy of Sciences, Bulgaria.}}

\vskip 10mm

\centerline{\sl Department of Physics} \centerline{\sl Ewha Womans
University} \centerline{\sl DaeHyun 11-1, Seoul 120-750, S. Korea}
\vspace*{0.6cm} \centerline{\tt ahn@ewha.ac.kr,
bozhilov@inrne.bas.bg}

\vskip 20mm

\baselineskip 18pt

\begin{center}
{\bf Abstract}
\end{center}

We develop an approach for solving the string equations of motion
and Virasoro constraints in any background which has some (unfixed)
number of commuting Killing vector fields. It is based on a specific
ansatz for the string embedding. We apply the above mentioned
approach for strings moving in $AdS_3\times S^3\times T^4$ with
2-form NS-NS B-field. We succeeded to find solutions for a large
class of string configurations on this background. In particular, we derive
dyonic giant magnon solutions in the $R_t \times S^3$ subspace,
and obtain the leading finite-size correction to the dispersion relation.

\end{titlepage}
\newpage
\baselineskip 18pt

\def\nn{\nonumber}
\def\tr{{\rm tr}\,}
\def\p{\partial}
\newcommand{\non}{\nonumber}
\newcommand{\bea}{\begin{eqnarray}}
\newcommand{\eea}{\end{eqnarray}}
\newcommand{\bde}{{\bf e}}
\renewcommand{\thefootnote}{\fnsymbol{footnote}}
\newcommand{\be}{\begin{eqnarray}}
\newcommand{\ee}{\end{eqnarray}}

\vskip 0cm

\renewcommand{\thefootnote}{\arabic{footnote}}
\setcounter{footnote}{0}

\setcounter{equation}{0}
\section{Introduction}

A very important development in the field of string theory
has been achieved for the case of AdS/CFT duality
\cite{AdS/CFT} between strings and conformal field theories in
various dimensions. The most developed case is the correspondence
between strings living in $AdS_5\times S^5$ and $\mathcal{N}=4$ SYM
in four dimensions. Another example is the duality
between strings on $AdS_4\times CP^3$ background and $\mathcal{N}=6$
super Chern-Simons-matter theory in three space-time dimensions. The
main achievements in the above examples are due to the discovery of
integrable structures on both sides of the correspondence. Many
other cases have been considered also \cite{RO}.
In particular, for the case of large coupling constant (large string tension),
relations between semi-classical limits of the string solutions and the
relevant objects in the dual gauge theories have been found, based on the
underlying integrable models appearing on both sides.

Another interesting area of research is the $AdS_3/CFT_2$
duality \cite{BSZ0912}-\cite{LS1312}, related to $AdS_3\times
S^3\times T^4$ and $AdS_3\times S^3\times S^3\times S^1$ string
theory backgrounds.
Here we will consider the classical string dynamics for the
relatively simpler case of $AdS_3\times S^3\times T^4$.

The paper is organized as follows. In Sec.2 we present our general
approach to string dynamics in curved backgrounds with B-field. In
Sec.3 we apply it to strings moving in $AdS_3\times S^3\times T^4$
with NS-NS B-field background. In Sec.4 we restrict ourselves to
giant magnon solutions \cite{HM06} and derive the dispersion relation,
including the leading finite-size effect on it.
Sec.5 is devoted to our concluding remarks.

\setcounter{equation}{0}
\section{Strings in curved backgrounds with B-field: the general approach}
Considering string dynamics in curved backgrounds with B-field,
we develop an approach, which will allow us to obtain exact string solutions in
sufficiently general string theory target spaces.

\subsection{Bosonic string action, equations of motion and constraints}

In our further considerations, we will use the Polyakov type action
for the bosonic string in a $D$-dimensional curved space-time with
metric tensor $g_{MN}(x)$, interacting with a background 2-form
gauge field $b_{MN}(x)$ via Wess-Zumino term \bea\nn &&S^{P}=\int
d^{2}\xi\mathcal{L}^P,\h \mathcal{L}^P
=-\frac{1}{2}\left(T\sqrt{-\gamma}\gamma^{mn}G_{mn}-Q\varepsilon^{mn}
B_{mn}\right),\\ \nn && \xi^m=(\xi^0,\xi^1)=(\tau,\sigma),\h m,n =
0,1,\eea where  \bea\nn &&G_{mn}= \p_m X^M\p_n X^N g_{MN},\h
B_{mn}=\p_{m}X^{M}\p_{n}X^{N} b_{MN}, \\ \nn &&(\p_m=\p/\p\xi^m,\h
M,N = 0,1,\ldots,D-1),\eea are the fields induced on the string
worldsheet, $\gamma$ is the determinant of the auxiliary worldsheet
metric $\gamma_{mn}$, and $\gamma^{mn}$ is its inverse. The position
of the string in the background space-time is given by
$x^M=X^M(\xi^m)$, and $T=1/2\pi\alpha'$, $Q$ are the string tension
and charge, respectively. If we consider the action $S^P$ as a
bosonic part of a supersymmetric one, we have to put $Q=\pm T$. In
what follows, $Q =T$.

The equations of motion for $X^M$ following from $S^P$ are: \bea \nn
&&-g_{LK}\left[\p_m\left(\sqrt{-\gamma}\gamma^{mn}\p_nX^K\right) +
\sqrt{-\gamma}\gamma^{mn}\Gamma^K_{MN}\p_m X^M\p_n X^N\right]\\
\label{em} &&=\frac{1}{2}H_{LMN}\epsilon^{mn}\p_m X^M\p_n X^N,\eea
where ($\p_M=\p/\p x^M$)
\bea\nn
&&\Gamma_{L,MN}=g_{LK}\Gamma^K_{MN}=\frac{1}{2}\left(\p_Mg_{NL}
+\p_Ng_{ML}-\p_Lg_{MN}\right),\\ \nn &&H_{LMN}= \p_L b_{MN}+ \p_M
b_{NL} + \p_N b_{LM},\eea are the components of the symmetric
connection corresponding to the metric $g_{MN}$, and the field
strength of the gauge field $b_{MN}$ respectively. The constraints
are obtained by varying the action $S^P$ with respect to
$\gamma_{mn}$: \bea\label{oc} \delta_{\gamma_{mn}}S^P=0\Rightarrow
\left(\gamma^{kl}\gamma^{mn}-2\gamma^{km}\gamma^{ln}\right)G_{mn}=0.\eea

\subsection{Gauge choice and ansatz}
In what follow we will  use {\it conformal gauge}
$\gamma^{mn}=\eta^{mn}=diag(-1,1)$ in which the string Lagrangian,
the Virasoro constraints and the equations of motion take the
following form: \bea\label{CG} &&\mathcal{L}
=\frac{T}{2}\left(G_{00}-G_{11}+2 B_{01}\right),\\ \nn
&&G_{00}+G_{11}=0,\h G_{01}=0, \\ \nn &&
g_{LK}\left[\left(\p_0^2-\p_1^2\right)X^K+ \Gamma_{MN}^K\left(\p_0
X^M \p_0 X^N-\p_1 X^M \p_1
X^N\right)\right]=H_{LMN}\p_0X^M\p_1X^N.\eea

Now, let us {\it suppose} that there exist some number
of commuting Killing vector fields along part of $X^M$ coordinates and split $X^M$ into two parts
\bea\nn
X^M=(X^\mu,X^a),\eea
where $X^\mu$ are the isometric coordinates,
while $X^a$ are the non-isometric ones.
The existence of isometric coordinates leads to the following conditions on the background fields:
\bea\label{cbf}
\p_\mu g_{MN}=0,\h \p_\mu b_{MN}=0.\eea
Then from the string action,
we can compute the conserved charges \bea\label{CC} Q_\mu=\int
d\sigma \frac{\p \mathcal{L}}{\p(\p_0 X^\mu)}\eea
under the above conditions.

Next, we introduce the following ansatz for the string embedding
\bea\label{A}
X^{\mu}(\tau,\sigma)=\Lambda^{\mu}\tau
+\tilde{X}^{\mu}(\alpha\sigma+\beta\tau),\h
X^{a}(\tau,\sigma)=\tilde{X}^{a}(\alpha\sigma+\beta\tau),\eea where
$\Lambda^{\mu}$, $\alpha$, $\beta$ are arbitrary parameters.
Further on, we will use the notation $\xi=\alpha\sigma+\beta\tau $.
Applying this ansatz, one can find that the equalities (\ref{CG}), (\ref{CC}) become
\bea\label{LA} \mathcal{L}=
\frac{T}{2}\Big[-(\alpha^2-\beta^2)g_{MN}\frac{d\tilde{X}^M}{d\xi}\frac{d\tilde{X}^N}{d\xi}
+2\Lambda^\mu\left(\beta g_{\mu N}+\alpha b_{\mu
N}\right)
\frac{d\tilde{X}^N}{d\xi}
+\Lambda^\mu\Lambda^\nu g_{\mu \nu}\Big],\eea

\bea\label{V1}
G_{00}+G_{11}=(\alpha^2+\beta^2)g_{MN}\frac{d\tilde{X}^M}{d\xi}\frac{d\tilde{X}^N}{d\xi}
+2\beta\Lambda^\mu g_{\mu N}
\frac{d\tilde{X}^N}{d\xi}
+\Lambda^\mu\Lambda^\nu g_{\mu \nu} =0,\eea

\bea\label{V2}  G_{01}&=&\alpha\beta
g_{MN}\frac{d\tilde{X}^M}{d\xi}\frac{d\tilde{X}^N}{d\xi}+\alpha\Lambda^\mu
g_{\mu N}\frac{d\tilde{X}^N}{d\xi} =0,\eea

\bea\nn &&-(\alpha^2-\beta^2)\left[g_{LK}\frac{d^2\tilde{X}^K}{d\xi^2}+
\Gamma_{L,MN}\frac{d\tilde{X}^M}{d\xi}\frac{d\tilde{X}^N}{d\xi}\right]
+2\beta\Lambda^\mu\Gamma_{L,\mu N}\frac{d\tilde{X}^N}{d\xi}
+\Lambda^\mu\Lambda^\nu \Gamma_{L,\mu\nu}
\\ \label{EM}
&&=
\alpha\Lambda^\mu H_{L\mu N}\frac{d\tilde{X}^N}{d\xi},\eea

\bea\label{Q} &&Q_{\mu}= \frac{T}{\alpha}\int d\xi\left[\left(\beta
g_{\mu N}+\alpha b_{\mu N}\right)\frac{d\tilde{X}^N}{d\xi}+\Lambda^\nu
g_{\mu \nu}\right].\eea

Our next task is to try to solve the equations of motion (\ref{EM})
for the isometric coordinates, i.e. for $L=\lambda$. Due
to the conditions (\ref{cbf}) imposed on the background fields, we
obtain that \bea\nn &&\Gamma_{\lambda,a b}=\frac{1}{2}\left(\p_a
g_{b \lambda}+\p_b
 g_{a \lambda}\right),\h  \Gamma_{\lambda,\mu a}=\frac{1}{2}\p_a g_{\mu\lambda}\h
 \Gamma_{\lambda,\mu \nu}=0,
 \\ \nn && H_{\lambda a b}=\p_a b_{b \lambda}+\p_b b_{\lambda a}, \h
 H_{\lambda \mu a}=\p_a b_{\lambda\mu}, \h  H_{\lambda \mu \nu}=0.
 \eea
By using this, one can find the following first integrals for
$\tilde{X}^\mu$: \bea\label{FIM} \frac{d\tilde{X}^{\mu}}{d\xi}=
\frac{1}{\alpha^2-\beta^2}\left[g^{\mu\nu}
\left(C_\nu-\alpha\Lambda^\rho
b_{\nu\rho}\right)+\beta\Lambda^\mu\right]
-g^{\mu\nu}g_{\nu a}\frac{d\tilde{X}^{a}}{d\xi},\eea
where $C_\nu$ are arbitrary integration constants.
Therefore, according to
our ansatz (\ref{A}), the solutions for the string coordinates
$X^\mu$ can be written as \bea\label{MS}
X^{\mu}(\tau,\sigma)=\Lambda^{\mu}\tau
+\frac{1}{\alpha^2-\beta^2}\int d\xi \left[g^{\mu\nu}
\left(C_\nu-\alpha\Lambda^\rho
b_{\nu\rho}\right)+\beta\Lambda^\mu\right]
-\int g^{\mu\nu}g_{\nu a}d\tilde{X}^{a}(\xi).\eea

Now, let us turn to the remaining equations of motion corresponding
to $L=a$, where \bea\nn &&\Gamma_{a,\mu b}=-\frac{1}{2}(\p_a g_{b
\mu}-\p_b g_{a \mu}),\h \Gamma_{a,\mu \nu}=-\frac{1}{2}\p_a
g_{\mu\nu},
\\ \nn &&H_{a \mu\nu}=\p_a b_{\mu\nu},\h H_{a \mu b}=-\p_a b_{b \mu}+\p_b b_{a \mu}.\eea
Taking this into account and replacing the first integrals for
$\tilde{X}^{\mu}$ already found, one can write
these equations in the form (prime is used for $d/d\xi$)
\bea\label{Ea} (\alpha^2-\beta^2)\left[h_{a
b}\tilde{X}^{b''}+\Gamma^{h}_{a,bc}\tilde{X}^{b'}\tilde{X}^{c'}\right] = 2\p_{[a}
A_{b]}\tilde{X}^{b'}-\p_a U, \eea where \bea\label{Ma} &&h_{a b}= g_{a
b}-g_{a \mu}g^{\mu\nu}g_{\nu b},\h
\Gamma^{h}_{a,bc}=\frac{1}{2}\left(\p_b h_{ca}+\p_c h_{ba}-\p_a
h_{bc}\right)
\\ \label{VPa} &&A_a= g_{a\mu}g^{\mu\nu}
\left(C_\nu-\alpha\Lambda^\rho
b_{\nu\rho}\right)+\alpha\Lambda^\mu b_{a\mu},
\\ \label{SP} &&U=
\frac{1/2}{\alpha^2-\beta^2}\left[\left(C_\mu-\alpha\Lambda^\rho
b_{\mu\rho}\right)g^{\mu\nu}
\left(C_\nu-\alpha\Lambda^\lambda
b_{\nu\lambda}\right)+\alpha^2\Lambda^\mu\Lambda^\nu
g_{\mu\nu}\right].\eea

One can show that the above equations for $\tilde{X}^{a}$ can be derived
from the effective Lagrangian \bea\nn \mathcal{L}^{eff}(\xi)
=\frac{1}{2}(\alpha^2-\beta^2) h_{a b}\tilde{X}^{a'}\tilde{X}^{b'}+A_a \tilde{X}^{a'}-U
.\eea The corresponding effective Hamiltonian is \bea\nn
\mathcal{H}^{eff}(\xi) =\frac{1}{2}(\alpha^2-\beta^2) h_{a
b}\tilde{X}^{a'}\tilde{X}^{b'}+U ,\eea
or in terms of the momenta $p_a$  conjugated to $\tilde{X}^a$
\bea\nn \mathcal{H}^{eff}(\xi)
=\frac{1}{2}(\alpha^2-\beta^2) h^{a
b}\left(p_a-A_a\right)\left(p_b-A_b\right)+U .\eea

The Virasoro constraints (\ref{V1}), (\ref{V2}) become:
\bea\label{V12} \frac{1}{2}(\alpha^2-\beta^2) h_{a
b}\tilde{X}^{a'}\tilde{X}^{b'}+U=0,\h
\alpha\Lambda^\mu C_\mu =0.\eea

Finally, let us write down the expressions for the conserved charges
(\ref{Q}) \bea\nn Q_\mu &=&\frac{T}{\alpha^2-\beta^2}\int d\xi
\left[\frac{\beta}{\alpha}C_\mu+\alpha\Lambda^\nu
g_{\mu\nu} + b_{\mu\nu}g^{\nu\rho}
\left(C_\rho-\alpha\Lambda^\lambda
b_{\rho\lambda}\right)\right. \\ \label{Qmu}
&+&\left.(\alpha^2-\beta^2) \left(b_{\mu
a}-b_{\mu\nu}g^{\nu\rho}g_{\rho a}\right)\tilde{X}^{a'}\right].\eea

\setcounter{equation}{0}
\section{Strings in $AdS_3\times S^3\times T^4$ with
NS-NS B-field}

The background geometry of this target space can be written in the following form
\footnote{The common radius $R$ of the three subspaces is set to 1, and
$q$ is the parameter used in \cite{HST1311}.}:
\bea\nn
&&ds^2_{AdS_3}=
-(1+r^2)dt^2+(1+r^2)^{-1}dr^2+r^2d\phi^2,\h
b_{t\phi}=q r^2,
\\ \nn &&ds^2_{S^3}=
d\theta^2+\sin^2\theta d\phi_1^2+\cos^2\theta d\phi_2^2,\h
b_{\phi_1\phi_2}=-q\cos^2\theta, \\ \nn
&&ds^2_{T^4}=(d\varphi^i)^2,\h i=1,2,3,4.\eea

Now, we apply the formulation given in Sec. 2.
According to our notations
\bea\nn
&&X^\mu=\left(t,\phi,\phi_1,\phi_2,\varphi^i\right),\h
X^a=\left(r,\theta\right),
\\ \nn &&g_{\mu\nu}=\left(g_{t t},g_{\phi
\phi},g_{\phi_1\phi_1},g_{\phi_2\phi_2},g_{ij}\right),\h
g_{ab}=\left(g_{r r},g_{\theta\theta}\right),\h g_{a \mu}=0,\h
h_{ab}=g_{ab},
\\ \nn &&b_{\mu\nu}=(b_{t \phi},b_{\phi_1\phi_2}),\h b_{a \nu}=0,
\\ \label{idc} &&A_a=0,\eea
where
\bea\nn &&g_{t t}=(g^{t t})^{-1}=-(1+r^2),\h
g_{\phi \phi}=(g^{\phi \phi})^{-1}=r^2,
\h g_{\phi_1 \phi_1}=(g^{\phi_1 \phi_1})^{-1}=\sin^2\theta,
\\ \nn &&g_{\phi_2 \phi_2}=(g^{\phi_2 \phi_2})^{-1}=\cos^2\theta,
\h g_{ij}=\left(g^{ij}\right)^{-1}= \delta _{ij},
\\ \nn
&&g_{r r}=(g^{r r})^{-1}=(1+r^2)^{-1},
\h g_{\theta\theta}=1,
\\ \label{bfs} &&b_{t\phi}=q r^2,\h b_{\phi_1\phi_2}=-q\cos^2\theta .\eea

Since $g_{a \mu}=0$, the solutions (\ref{MS}) for the coordinates
$X^\mu$ are simplified to
\bea\label{Xmu}
X^{\mu}(\tau,\sigma)=\Lambda^{\mu}\tau +\tilde{X}^\mu(\xi)
=\Lambda^{\mu}\tau
+\frac{1}{\alpha^2-\beta^2}\int d\xi \left[g^{\mu\nu}
\left(C_\nu-\alpha\Lambda^\rho
b_{\nu\rho}\right)+\beta\Lambda^\mu\right]
,\eea where $g^{\mu\nu}$ and $b_{\nu\rho}$ must be replaced from
above.

Now, we want to find the solutions for the non-isometric string coordinates $X^a$.
To this end we have to solve the equations (\ref{Ea}), which in the
case at hand reduce to \bea\label{Eqa} (\alpha^2-\beta^2)\left[g_{a
b}\tilde{X}^{b''}+\Gamma_{a,bc}\tilde{X}^{b'}\tilde{X}^{c'}\right] +\p_a \sum_{b=r,\theta} U_b=0,\eea
where the scalar potential $U$ in (\ref{SP}) is represented as a sum of two
parts: $ U_r= U_r(r)$ for the $AdS_3$ subspace and $ U_\theta= U_\theta(\theta)$
for the $S^3$ subspace of the background.

Taking into account that the metric $g_{a b}$ is diagonal, one can
find the following two first integrals of (\ref{Eqa})
\bea\label{FIa} \tilde{X}^{a'}=
\sqrt{\frac{C_a-2U_a}{(\alpha^2-\beta^2)g_{aa}}}.\eea
It follows from here that \bea\label{xi} d\xi=
\frac{d\tilde{X}^a}{\sqrt{\frac{C_a-2U_a}{(\alpha^2-\beta^2)g_{aa}}}}.\eea
So, we have two different expressions for $ d\xi$, which obviously
must coincide. This is a condition for self-consistency. It leads to
\bea\label{O} \int\frac{dr}{\sqrt{\frac{C_r-2U_r}{g_{rr}}}}=
\int\frac{d\theta}{\sqrt{\frac{C_\theta-2U_\theta}{g_{\theta\theta}}}},\eea
which actually gives implicitly the ``orbit'' $r(\theta)$, i.e. how the radial coordinate
$r$ on $AdS_3$ depends on the angle $\theta$ in $S^3$.

Now, we have to check if the first integrals for $\tilde{X}^a(\xi)$ are
compatible with the Virasoro constraints (\ref{V12}). Replacing
$\tilde{X}^{a'}$ in the first of them, one finds \bea\nn C_r+C_\theta=0.\eea

Thus, we found all first integrals of the string equations of
motion, compatible with the Virasoro constraints, which reduce to
algebraic relations between the embedding parameters and the
integration constants.

Now, let us give the expressions for the conserved charges
(\ref{Qmu}), corresponding to the isometric coordinates. \bea\label{EsS}
&&-Q_t\equiv E_s=\frac{T}{\alpha^2-\beta^2}\left[\left(\alpha
\Lambda^t-\frac{\beta}{\alpha}C_t-q \ C_\phi\right)\int
d\xi+\alpha (1-q^2)\Lambda^t\int d\xi r^2\right],
\\ \nn &&Q_\phi\equiv
S=\frac{T}{\alpha^2-\beta^2}\Bigg[\left(\frac{\beta}{\alpha}C_\phi+q C_t+
q^2 \alpha \Lambda^\phi\right)\int
d\xi +(1-q^{2})\alpha\Lambda^{\phi}\int
d\xi r^2
\\ \label{Sf} &&-\left(q C_t+q^2\alpha
\Lambda^\phi\right)\int
\frac{d\xi}{1+r^2}\Bigg], \eea
\bea\label{Qt} &&Q_{\phi_1}\equiv
J_{1}=\frac{T}{\alpha^2-\beta^2}\Bigg[\left(\frac{\beta}{\alpha}C_{\phi_1}+
\alpha\Lambda^{\phi_1}-q C_{\phi_2}\right)\int d\xi
-(1-q^2)\alpha
\Lambda^{\phi_1}\int
\cos^2\theta d\xi\Bigg],
\\ \nn &&Q_{\phi_2}\equiv
J_{2}=\frac{T}{\alpha^2-\beta^2}\Bigg[\left(\frac{\beta}{\alpha}C_{\phi_2}
-q\left(C_{\phi_1}+q\alpha\Lambda^{\phi_2}\right)\right)\int d\xi \\
\nn &&+(1-q^2)\alpha\Lambda^{\phi_2}\int \cos^2\theta d\xi
+q\left(C_{\phi_1}+q\alpha\Lambda^{\phi_2}\right)\int
\frac{d\xi}{1-\cos^2\theta}\Bigg],\eea
\bea \label{Ji} &&Q_{i}\equiv J_i^T= \frac{T}{\alpha^2-\beta^2}
\left(\frac{\beta}{\alpha}C_i +\alpha
\Lambda^j \delta_{ij}\right)\int d\xi.\eea
Here we used the following notations:
$E_s$ is the string energy, $S$ is the spin in $AdS_3$, $J_{1}$ and
$J_{2}$ are the two angular momenta in $S^3$, while $J_i^T$ are the four angular momenta on $T^4$.

The explicit expressions for the string coordinates, the ``orbit'' $r(\theta)$,
and the conserved charges in this background are given in Appendix A.

\setcounter{equation}{0}
\section{Giant magnon solutions}

The giant magnon string solution was found in \cite{HM06}.
It is a specific string configuration, living in the $R_t\times S^2$ subspace of $AdS_5\times S^5$ with an angular momentum $J_1$ which goes to $\infty$.
A similar configuration, dyonic giant magnon, has been obtained in \cite{CDO06} which moves in $R_t\times S^3$ subspace with two angular momenta 
$J_1,\ J_2$ with $J_1\to\infty$. These classical configurations have played an important role in understanding exact, quantum aspects of the AdS/CFT correspondence.
In particular, corrections due to a large but finite $J_1$ obtained in \cite{AFZ06} and \cite{HS08} can provide a nontrivial check for the exact worldsheet
$S$-matrix.

In this section we provide similar string solutions in $AdS_3\times S^3\times T^4$ with NS-NS
B-field for a large but finite $J_1$. A giant magnon solution with infinite angular momentum has been constructed in a recent paper \cite{HST1311}  
with a dispersion relation
\footnote{The terms proportional to $q$ are due to the nonzero B-field on $S^3$.}
\bea\label{Tdr} E_s-J_1= \sqrt{(J_2-q T \Delta\phi_1)^2 +4 T^2(1-q^2) \sin^2\frac{\Delta\phi_1}{2}}.\eea
This relation is already quite different from those for the ordinary (dyonic) giant magnons. 
We will show that there exist even bigger differences for the finite-size corrections.

\subsection{Exact results}

In order to consider dyonic giant magnon solutions, we restrict our general ansatz (\ref{A}) in
the following way:
\bea\nn &&X^t\equiv t=\kappa\tau,\h \mbox{i.e.}\h
\Lambda^t=\kappa,\h \tilde{X}^t(\xi)=0,
\\ \nn &&X^\phi\equiv \phi=0,\h \mbox{i.e.}\h
\Lambda^\phi=0,\h \tilde{X}^\phi(\xi)=0
\\ \nn &&X^r \equiv r=\tilde{X}^r(\xi)=0,
\\ \nn &&X^{\phi_1}\equiv\phi_1=\omega_1\tau
+\tilde{X}^{\phi_1}(\xi),\h \mbox{i.e.}\h
\Lambda^{\phi_1}=\omega_1,
\\ \nn &&X^{\phi_2}\equiv\phi_2=\omega_2\tau
+\tilde{X}^{\phi_2}(\xi),\h \mbox{i.e.}\h \Lambda^{\phi_2}=\omega_2,
\\ \nn &&X^\theta \equiv \theta =\tilde{X}^\theta(\xi),\h X^{\varphi^i}\equiv\varphi^i=0.\eea

As a result, we can claim that \bea\nn C_t=\beta\kappa ,\eea
which comes from $\frac{d\tilde{X}^{t}}{d\xi}=0$.

Now, we can rewrite the first
integrals for $\tilde{X}^\mu$ on $S^3$ as
\bea\label{FmuR}
 &&\frac{d\tilde{X}^{{\phi_1}}}{d\xi}=\frac{1}{\alpha^2-\beta^2}
\left[\left(C_{\phi_1}+q\alpha\omega_2\right)\frac{1}{1-\chi}
+\beta\omega_1-q\alpha\omega_2\right],
\\ \nn &&\frac{d\tilde{X}^{{\phi_2}}}{d\xi}=\frac{1}{\alpha^2-\beta^2}\left(\frac{C_{\phi_2}}{\chi}
+\beta\omega_2-q\alpha\omega_1\right),\eea where
$\chi=\cos^2\theta$.

The first Virasoro constraint, which in the case under consideration
is the first integral of the equation of motion for $\theta$, reduces to
\bea\label{cp2R} \left(\frac{d\chi}{d\xi}\right)^2 &=&
\frac{4}{(\alpha^2-\beta^2)^2}\chi(1-\chi)
\Bigg[(\alpha^2+\beta^2)\kappa^2-\frac{\left(C_{\phi_1}+q\alpha\omega_2\chi\right)^2}{1-\chi}-\frac{\left(C_{\phi_2}
-q\alpha\omega_1\chi\right)^2}{\chi}
\\ \nn &&
-\alpha^2(\omega_2^2-\omega_1^2)\chi-\alpha^2\omega_1^2\Bigg].\eea
Also, the second Virasoro constraint becomes \bea\label{V2R} \omega_1 C_{\phi_1}+\omega_2
C_{\phi_2}+\beta\kappa^2=0.\eea

Taking (\ref{V2R}) into account, we can rewrite (\ref{cp2R}) as
\bea\label{V1f}  \left(\frac{d\chi}{d\xi}\right)^2=4(1-q^{2}) \frac{\omega_1^2}{\alpha^2}
\frac{1-u^2}{(1-v^2)^2}(\chi_p-\chi)(\chi-\chi_m)(\chi-\chi_n),\eea
where \bea\label{hes} &&\chi_p+\chi_m+\chi_n =
\frac{-\left(v^2W+\left(W+u^2-2+q^2)\right)
+2q\left(uvW+K(1-u^2)\right)\right)}{(1-q^{2})(1-u^2)},
 \\ \nn
&&\chi_p\chi_m+\chi_p\chi_n+\chi_m\chi_n= -\frac{\left(1+v^2\right)W+K^2-\left(v W-u K\right)^2
-1+2q K}{(1-q^{2})(1-u^2)},
\\ \nn &&\chi_p\chi_m\chi_n= -\frac{K^2}{(1-q^2) (1-u^2)},\eea and
we introduced the notations
\bea\nn v=-\frac{\beta}{\alpha},\h
u=\frac{\omega_2}{\omega_1},\h
W=\left(\frac{\kappa}{\omega_1}\right)^2,\h
K=\frac{C_{\phi_2}}{\alpha\omega_1}.\eea
This leads to
\bea\label{GMsdxi}
d\xi=\frac{\alpha}{2\omega_1}\frac{1-v^2}{\sqrt{(1-q^{2})(1-u^2)}}
\frac{d\chi}{\sqrt{(\chi_p-\chi)(\chi-\chi_m)(\chi-\chi_n)}}.\eea

Integrating (\ref{GMsdxi}) and inverting $\xi(\chi)$ to $\chi(\xi)\equiv \cos^2[\theta(\xi)]$,
one finds the following explicit solution
\bea\label{cossol} \chi = (\chi_p-\chi_n)\ \mathbf{DN}^2\left[\frac{\sqrt{(1-q^{2})(1-u^2)(\chi_p-\chi_n)}}{1-v^2}
\ \omega_1(\sigma-v \tau),\frac{\chi_p-\chi_m}{\chi_p-\chi_n}\right]+\chi_n ,\eea
where $\mathbf{DN}$ is one of the Jacobi elliptic functions.

Next, we integrate (\ref{FmuR}), and according to our ansatz,
obtain that the solutions for the isometric angles on $S^3$ are given by
\bea\label{1one} &&\phi_1= \omega_1 \tau+ \frac{2}{\sqrt{(1-q^{2})(1-u^2)(\chi_p-\chi_n)}}
\\ \nn &&\left[\frac{v W-K u+q u}{1-\chi_p}\ \mathbf{\Pi}\left(\arcsin\sqrt{\frac{\chi_p-\chi}
{\chi_p-\chi_m}},-\frac{\chi_p-\chi_m}{1-\chi_p},\frac{\chi_p-\chi_m}{\chi_p-\chi_n}\right)\right.
\\ \nn &&\left.
-(v+q u)\mathbf{F}\left(\arcsin\sqrt{\frac{\chi_p-\chi}
{\chi_p-\chi_m}},\frac{\chi_p-\chi_m}{\chi_p-\chi_n}\right)\right]\eea
\bea\label{Isols} &&\phi_2=\omega_2 \tau+\frac{2}{\sqrt{(1-q^{2})(1-u^2)(\chi_p-\chi_n)}}
\\ \nn &&\left[\frac{K}{\chi_p}\ \mathbf{\Pi}\left(\arcsin\sqrt{\frac{\chi_p-\chi}
{\chi_p-\chi_m}},1-\frac{\chi_m}{\chi_p},\frac{\chi_p-\chi_m}{\chi_p-\chi_n}\right)\right.
\\ \nn &&\left.
-(u v+q)\mathbf{F}\left(\arcsin\sqrt{\frac{\chi_p-\chi}
{\chi_p-\chi_m}},\frac{\chi_p-\chi_m}{\chi_p-\chi_n}\right)\right],\eea
where $\mathbf{F}$ and $\mathbf{\Pi}$ are the incomplete elliptic integrals of first and third kind.

By using (\ref{GMsdxi}), one can find also the conserved
quantities, namely, the string energy $E_s$ and the two angular
momenta $J_1$, $J_2$ : \bea\label{GMsEs} E_s=
2T\frac{(1-v^2)\sqrt{W}}{\sqrt{(1-q^{2})(1-u^2)(\chi_p-\chi_n)}}\
\mathbf{K}(1-\epsilon),\eea

\bea\nn &&J_1= \frac{2T}{\sqrt{(1-q^{2})(1-u^2)(\chi_p-\chi_n)}}
\Big\{\left[1-v^2 W+K(uv-q)\right]\mathbf{K}(1-\epsilon)
\\ \label{Jtf}
&&-(1-q^2)\left[\chi_n \ \mathbf{K}(1-\epsilon)+(\chi_p-\chi_n)\ \mathbf{E}(1-\epsilon)\right]\Big\},\eea

\bea\nn &&J_2= \frac{2T}{\sqrt{(1-q^{2})(1-u^2)(\chi_p-\chi_n)}}
\Big\{(1-q^{2}) u\left[\chi_n \ \mathbf{K}(1-\epsilon)+(\chi_p-\chi_n)\ \mathbf{E}(1-\epsilon)\right]
\\ \label{Jf} &&-\left[K v+q\left(v W-K u\right)+q^{2}u\right]\ \mathbf{K}(1-\epsilon)
\\ \nn &&+q\frac{v W-K u+q u}{1-\chi_p}
\mathbf{\Pi}\left(-\frac{\chi_p-\chi_n}{1-\chi_p}(1-\epsilon),1-\epsilon\right)
\Big\},\eea
where $\mathbf{K}$, $\mathbf{E}$ and $\mathbf{\Pi}$, are the complete elliptic integrals
of first, second and third kind, and $\epsilon$ is defined as
\bea\label{eps}
\epsilon=\frac{\chi_m-\chi_n}{\chi_p-\chi_n}.\eea

We will need also the expression for the angular difference $\Delta\phi_1$.
It can be found to be
\bea\label{dtt} &&\Delta\phi_1 = \frac{2}{\sqrt{(1-q^{2})(1-u^2)(\chi_p-\chi_n)}}
\\ \nn &&\left[\frac{v W-K u+q u}{1-\chi_p}\
\mathbf{\Pi}\left(-\frac{\chi_p-\chi_n}{1-\chi_p}(1-\epsilon),1-\epsilon\right)
-(v+q u)\ \mathbf{K}\left(1-\epsilon\right)\right].\eea

The expressions (\ref{GMsEs}), (\ref{Jtf}), (\ref{Jf}), (\ref{dtt}) are for the
finite-size dyonic strings living in the $R_t\times S^3$ subspace of
$AdS_3\times S^3\times T^4$.

\subsection{Leading finite-size effect on the dispersion relation}

In order to find the leading finite-size effect on the dispersion relation,
we have to consider the limit $\epsilon\to 0$, since
$\epsilon= 0$ corresponds to the infinite-size case. In this subsection we restrict ourselves to
the particular case when $\chi_n = K=0$\footnote{As we will see later on, this choice allow us
to reproduce the dispersion relation in the infinite volume limit \cite{HST1311}.}.
Then the third equation in (\ref{hes}) is satisfied identically, while the other two simplify to
\bea\label{sce} &&\chi_p+\chi_m =
\frac{2-(1+v^2)W-u^2-2q(u v W+\frac{q}{2})}{(1-q^{2})(1-u^2)},
\\ \nn
&&\chi_p\chi_m= \frac{(1-W)(1-v^2W)}{(1-q^{2})(1-u^2)},\eea
and $\epsilon$ becomes
\bea\label{epss}
\epsilon=\frac{\chi_m}{\chi_p}.\eea
The relevant expansions of the
parameters are \bea\nn
&&\chi_p=\chi_{p0}+\left(\chi_{p1}+\chi_{p2}\log(\epsilon)\right)\epsilon,
\h W=1+W_{1}\epsilon ,
\\
\label{Dpars} &&v=v_0+\left(v_1+v_2\log(\epsilon)\right)\epsilon,
\h u=u_0+\left(u_1+u_2\log(\epsilon)\right)\epsilon.\eea
Replacing (\ref{epss}), (\ref{Dpars}) into (\ref{sce}), one finds the following solutions
in the small $\epsilon$ limit
\bea\label{4sol} &&\chi_{p0}=
\frac{1-v^2_0-u^2_0-2q(u_0 v_0+\frac{q}{2})}{(1-q^{2})(1-u_0^2)},
\\ \nn &&\chi_{p1}= -\frac{v_0+q u_0}{(1-q^{2})^2(1-v_0^2)(1-u_0^2)^2}\times
\\ \nn &&\Big[\Big(1-v_0^2-u_0^2-2q(u_0v_0+\frac{q}{2})\Big)\Big(v_0^3
+q u_0(1+3v_0^2)-v_0(1-2u_0^2-2q^2)\Big)
\\ \nn &&+2(1-q^{2})(1-v_0^2)\left((1-u_0^2)v_1+(u_0v_0+q)u_1\right)\Big]
\\ \nn &&\chi_{p2}= -\frac{2(v_0+q u_0)\left((1-u_0^2)v_2+(u_0 v_0+q)u_2\right)}
{(1-q^{2})(1-u_0^2)^2}
\\ \nn &&W_1= -\frac{\left(1-v^2_0-u^2_0-2q(u_0 v_0+\frac{q}{2})\right)^2}
{(1-q^{2})(1-u_0^2)(1-v_0^2)}.\eea

The coefficients in the expansions of $v$ and $u$, will be obtained by imposing the conditions that
$J_2$ and $\Delta\phi_1$ do not depend on $\epsilon$,
as in the cases without $B$-field ($AdS_5\times S^5$ and $AdS_4\times CP^3$)
and their $TsT$-deformations, where the $B$-field is nonzero, but its contribution is different.

Expanding (\ref{Jf}) and (\ref{dtt}) to the leading order in $\epsilon$ (now $\chi_n = K =0$),
one finds that on the solutions (\ref{4sol})
\bea\label{J} &&J_2=2T\left(\frac{u_0 \sqrt{1-u_0^2-v_0^2-2q(u_0v_0+\frac{q}{2})}}{1-u_0^2}\right.
\\ \nn &&\left.+q\arcsin\sqrt{\frac{1-u_0^2-v_0^2-2q(u_0v_0+\frac{q}{2})}{(1-q^2)(1-u_0^2)}}\right),\eea

\bea\label{Dtt} \Delta\phi_1=2\arcsin\sqrt{\frac{1-u_0^2-v_0^2-2q(u_0v_0+\frac{q}{2})}{(1-q^2)(1-u_0^2)}} ,\eea
\bea\label{u1} &&u_1= \frac{1-u_0^2-v_0^2-2q(u_0v_0+\frac{q}{2})}{4(1-q^2)(1-u_0^2)}
\\ \nn &&\times
\left[u_0\left(1-\log 16 -v_0^2(1+\log 16)\right)-2q v_0\log 16)\right],\eea
\bea\label{v1} &&v_1 = \frac{1-u_0^2-v_0^2-2q(u_0v_0+\frac{q}{2})}{4(1-q^2)(1-u_0^2)(1-v_0^2)}
\\ \nn &&\times
\left[v_0\left((1-4q^2)(1-\log 16)-u_0^2(5-\log 4096)\right)\right.
\\ \nn && \left.-v_0^3\left(1-\log 16-u_0^2(1+\log 16)\right)
-4q u_0\left(1-\log 4+v_0^2(1-\log 64)\right)\right],\eea
\bea\label{u2} &&u_2 = \frac{\left(u_0(1+v_0^2)+2q v_0\right)
\left(1-u_0^2-v_0^2-2q(u_0v_0+\frac{q}{2})\right)}{4 (1-q^2)(1-v_0^2)},\eea
\bea\label{v2} &&v_2 = \frac{1-u_0^2-v_0^2-2q(u_0v_0+\frac{q}{2})}{4(1-q^2)(1-u_0^2)(1-v_0^2)}
\\ \nn &&\times
\left[v_0\left(1-v_0^2-u_0^2(3+v_0^2)\right)
-2q\left(u_0(1+3v_0^2)+2q v_0\right)\right].\eea

Now, let us turn to the energy-charge relation.
Expanding (\ref{GMsEs}) and (\ref{Jtf}) in $\epsilon$ and taking into account the solutions
(\ref{4sol}), (\ref{u1}) -(\ref{v2}), we obtain
\bea\label{EsJtp} E_s-J_1= 2T\frac{\sqrt{1-u_0^2-v_0^2-2q(u_0v_0+\frac{q}{2})}}{1-u_0^2}
\left(1-\frac{1-u_0^2-v_0^2-2q(u_0v_0+\frac{q}{2})}{4(1-q^2)}\ \epsilon\right).\eea
The expression for $\epsilon$ can be found from the expansion of $J_1$.
To the leading order, it is given by
\bea\label{seps} \epsilon= 16 \exp\left[-\frac{J_1}{T}
\frac{\sqrt{1-u_0^2-v_0^2-2q(u_0v_0+\frac{q}{2})}}{1-v_0^2}
-2\frac{1-u_0^2-v_0^2-2q(u_0v_0+\frac{q}{2})}{(1-v_0^2)(1-u_0^2)}\right].\eea

Next, we would like to express the right hand side of (\ref{EsJtp}) in terms of $J_2$ and $\Delta\phi_1$.
To this end, we solve (\ref{J}), (\ref{Dtt}) with respect to $u_0$, $v_0$. The result is
\bea\label{u0} && u_0= \frac{J_2-q T \Delta\phi_1}
{\sqrt{(J_2-q T \Delta\phi_1)^2+4(1-q^2) T^2 \sin^2\frac{\Delta\phi_1}{2}}},
\\ \label{v0} && v_0= \frac{T(1-q^2)\sin \Delta\phi_1-q(J-q T \Delta\phi_1)}
{\sqrt{(J_2-q T \Delta\phi_1)^2+4(1-q^2) T^2 \sin^2\frac{\Delta\phi_1}{2}}}.\eea

Replacing (\ref{u0}), (\ref{v0}) into (\ref{EsJtp}), (\ref{seps}), one finds
\bea\label{EsJtpf} &&E_s-J_1=\sqrt{(J_2-q T \Delta\phi_1)^2+4(1-q^2) T^2 \sin^2\frac{\Delta\phi_1}{2}}
\\ \nn
&&\left(1-\frac{(1-q^2) T^2 \sin^4\frac{\Delta\phi_1}{2}}{(J_2-q T \Delta\phi_1)^2+4(1-q^2)
T^2 \sin^2\frac{\Delta\phi_1}{2}}\ \epsilon\right),\eea
where
\bea\label{sepsf} \epsilon=16\ e^{-\frac{2\left(J_1+\sqrt{(J_2-q T \Delta\phi_1)^2+4(1-q^2) T^2 \sin^2\frac{\Delta\phi_1}{2}}\right)
\sqrt{(J_2-q T \Delta\phi_1)^2+4(1-q^2) T^2 \sin^2\frac{\Delta\phi_1}{2}}\sin^2\frac{\Delta\phi_1}{2}}
{(J_2-q T \Delta\phi_1)^2+4T^2\sin^4\frac{\Delta\phi_1}{2}
+2q T\sin\Delta\phi_1\left((J_2-q T \Delta\phi_1)+\frac{q}{2} T\sin\Delta\phi_1\right)}}.\eea

Our result matches with that of  \cite{HST1311} in (\ref{Tdr}) whebn we take $\epsilon\to 0$ limit by sending $J_1\to\infty$.
This dispersion relation is different from the ordinary giant magnon which are

with the infinite-size dispersion relation,
found in.
By setting , one finds that the two results coincide
\footnote{It was explained in \cite{HST1311} why the angular difference $\Delta\phi_1$
must be identified with the momentum $p$ of the magnons in the dual spin chain.}

The dispersion relation for the ordinary giant magnon with
one nonzero angular momentum cam be obtained by setting $J_2=1$ and taking the limit
$T\to\infty$. To take into account the {\it leading} finite-size effect only,
we restrict ourselves to the case when $\frac{J_1}{T}>>1$.
The result is the following:
\bea\label{1spin} E_s-J_1=T \sqrt{p^2 q^2+4(1-q^2)\sin^2 \frac{p}{2}}\left(
1-\frac{\left(1-q^2\right)\sin^4\frac{p}{2}}{p^2 q^2+4(1-q^2)\sin^2 \frac{p}{2}}\ \epsilon\right),\eea
where
\bea\nn &&\epsilon =16 \exp\left[\frac{-2}{q^2(p-\sin p)^2+4 \sin^4\frac{p}{2}}
\left(\frac{J_1}{T}+\sqrt{p^2 q^2+4(1-q^2)\sin^2 \frac{p}{2}}\right)\right.
\\ \nn && \left.\sqrt{p^2 q^2+4(1-q^2)\sin^2 \frac{p}{2}}\  \sin^2 \frac{p}{2}\right].\eea

\setcounter{equation}{0}
\section{Concluding Remarks}
Here, we presented an approach to string dynamics in curved
backgrounds with nonzero 2-form B-field, which allows us to find the
first integrals for the string coordinates along the isometric
directions of the background and the corresponding conserved
charges. This leads to dimensional reduction of the problem. It
remains to solve the equations of motion for the non-isometric
string coordinates and the Virasoro constraints. This can be done
for fixed background fields. As an example we have considered
string dynamics on $AdS_3\times S^3\times T^4$. We succeeded to
find all solutions of the string equations of motion for this
case, and to reduce the Virasoro constraints to algebraic relations
among the embedding parameters and the integration constants. The
resulting family of string configurations may have very different
properties for different values of the parameters involved.
That is why, we concentrated on the finite-size dyonic giant magnon solutions in this background.
We have shown that the finite-size dispersion relation of (dyonic) giant magnon solution
in this background is different from the analogous
ones in $AdS_5\times S^5$, $AdS_4\times CP^3$ and their $\gamma$-deformations.

Our results on the leading finite-size correction to the dispersion relation
can provide an important check for the exact integrability conjecture and $S$-matrix elements based on it. 
We will report on this soon.
Another possible direction of further investigation is to consider strings
moving in $AdS_3\times S^3\times S^3\times S^1$ which has smaller set of isometric coordinates, hence, needs to solve more nontrivial
equations of motion.

\section*{Acknowledgements}
We would like to thank A. A. Tseytlin for providing the useful information.
This work was supported in part by the WCU Grant No. R32-2008-000-101300, the Research fund no. 1-2008-2935-001-2
 by Ewha Womans University (CA), and the Brain Pool program
(131S-1-3-0534) from  the MSIP Ministry of Science, Ict and future
Planning. 

\def\theequation{A.\arabic{equation}}
\setcounter{equation}{0}
\begin{appendix}

\section{Explicit exact solutions in $AdS_3\times S^3\times T^4$ with
NS-NS B-field}

Let start with the solutions for the string coordinates in $AdS_3$ subspace.
By using (\ref{SP}),
 (\ref{idc}) and (\ref{bfs}), one can find that the scalar potential $U_r$ in  (\ref{Eqa}) is given by
\bea\label{ur} &&U_r(r)=\frac{1}{2(\alpha^2-\beta^2)}\Bigg[\left(\alpha \Lambda^\phi\right)^2  r^2  -
\left(\alpha \Lambda^t\right)^2(1+r^2)
\\ \nn &&+ \frac{\left(C_\phi +q\alpha \Lambda^t r^2\right)^2}{r^2}-
 \frac{\left(C_t- q\alpha \Lambda^\phi r^2\right)^2}{1+r^2} \Bigg] .\eea

After introducing the variable \bea\label{yr} y=r^2,\eea and replacing (\ref{ur}) into (\ref{xi})
one can rewrite it in the following form
\bea\label{dxidy} d\xi=\frac{\alpha^2-\beta^2}{2\alpha
\sqrt{(1-q^2)\left[\left(\Lambda^\phi\right)^2-\left(\Lambda^t\right)^2\right]}}
\frac{dy}{\sqrt{(y_p-y)(y-y_m)(y-y_n)}} ,\eea where
\bea\nn 0\leq y_m<y<y_p,\h y_n<0,\eea
and $y_p$, $y_m$, $y_n$ satisfy the relations
\bea\nn &&y_p+y_m+y_n=\frac{1}{\alpha^{2}(1-q^2)\left[\left(\Lambda^\phi\right)^2-\left(\Lambda^t\right)^2\right]}
\\ \nn
&&\left[C_r(\alpha^2-\beta^2)-\alpha\left(\alpha\left(\Lambda^\phi\right)^2-2\alpha\left(\Lambda^t\right)^2\right)
+2q\left(C_\phi\Lambda^t+C_t\Lambda^\phi\right)+q^2\alpha\left(\Lambda^t\right)^2\right],
\\ \label{ypm} && y_p y_m+y_p y_n+y_m y_n = -\frac{1}{\alpha^{2}(1-q^2)\left[\left(\Lambda^\phi\right)^2-\left(\Lambda^t\right)^2\right]}
\\ \nn &&\left[C_r(\alpha^2-\beta^2)+C_t^2-C_\phi^2+\alpha^2\left(\Lambda^t\right)^2-2q\alpha C_\phi\Lambda^t\right],
\\ \nn && y_p y_m y_n=- \frac{C_\phi^2}{\alpha^{2}(1-q^2)\left[\left(\Lambda^\phi\right)^2-\left(\Lambda^t\right)^2\right]}.\eea

Integrating (\ref{dxidy}) and inverting
\bea\nn \xi(y)= \frac{\alpha^2-\beta^2}{\alpha
\sqrt{(1-q^2)\left[\left(\Lambda^\phi\right)^2-\left(\Lambda^t\right)^2\right](y_p-y_n)}}\
\mathbf{F}\left(\arcsin\sqrt{\frac{y_p-y}{y_p-y_m}},\frac{y_p-y_m}{y_p-y_n}\right)\eea
to $y(\xi)$, one finds the following solution
\bea\label{yxi} y(\xi)=(y_p-y_n)\ {\mathbf{DN}}^2\left[\frac{\alpha
\sqrt{(1-q^2)\left[\left(\Lambda^\phi\right)^2-\left(\Lambda^t\right)^2\right](y_p-y_n)}}
{\alpha^2-\beta^2}\ \xi,\frac{y_p-y_m}{y_p-y_n}\right]+y_n,\eea
where $\mathbf{F}$ is the incomplete elliptic integral of first kind and
$\mathbf{DN}$ is one of the Jacobi elliptic functions.

Next, we will compute $\tilde{X}^t(\xi)$ and $\tilde{X}^\phi(\xi)$ entering (\ref{Xmu}).
Integrating
\bea\nn &&\frac{d\tilde{X}^t}{d\xi}= \frac{1}{\alpha^2-\beta^2}
\left[\beta\Lambda^t +q\alpha\Lambda^\phi
-\left(C_t+q\alpha\Lambda^\phi\right)\frac{1}{1+y}\right],
\\ \nn &&\frac{d\tilde{X}^\phi}{d\xi}= \frac{1}{\alpha^2-\beta^2}
\left(\beta\Lambda^\phi+q\alpha\Lambda^t+\frac{C_\phi}{y}\right),\eea
and using (\ref{yxi}), we obtain the following solutions for the string coordinates $t$, $\phi$,
in accordance with our ansatz
\bea\label{tsol} &&t(\tau,\sigma)=\Lambda^t\tau+\frac{1}{\alpha\sqrt{(1-q^2)\left[\left(\Lambda^\phi\right)^2
-\left(\Lambda^t\right)^2\right](y_p-y_n)}}
\\ \nn &&\left[\left(\beta \Lambda^t+q \alpha\Lambda^\phi\right)\
\mathbf{F}\left(\arcsin\sqrt{\frac{y_p-y}{y_p-y_m}},\frac{y_p-y_m}{y_p-y_n}\right)\right.
\\ \nn &&-\left. \frac{C_t+q \alpha\Lambda^\phi}{1+y_p}
\ \mathbf{\Pi}\left(\arcsin\sqrt{\frac{y_p-y}{y_p-y_m}},
\frac{y_p-y_m}{1+y_p},\frac{y_p-y_m}{y_p-y_n}\right)\right]
\\ \label{Ff} &&\phi(\tau,\sigma) =\Lambda^\phi\tau + \frac{1}{\alpha\sqrt{(1-q^2)\left[\left(\Lambda^\phi\right)^2
-\left(\Lambda^t\right)^2\right](y_p-y_n)}}
\\ \nn &&\left[\left(\beta \Lambda^\phi+q\alpha\Lambda^t\right)\
\mathbf{F}\left(\arcsin\sqrt{\frac{y_p-y}{y_p-y_m}},\frac{y_p-y_m}{y_p-y_n}\right)\right.
\\ \nn &&+\left. \frac{C_\phi}{y_p}
\ \mathbf{\Pi}\left(\arcsin\sqrt{\frac{y_p-y}{y_p-y_m}},
\frac{y_p-y_m}{y_p},\frac{y_p-y_m}{y_p-y_n}\right)\right],\eea
where $\mathbf{\Pi}$ is the incomplete elliptic integral of third kind.

Let us compute now the string energy and spin on the solutions found.
Starting from (\ref{EsS}), (\ref{Sf}), we obtain
\bea\label{Ess} &&E_s= \frac{2 T}{\sqrt{(1-q^2)\left[\left(\Lambda^\phi\right)^2
-\left(\Lambda^t\right)^2\right](y_p-y_n)}}
\\ \nn &&\left[\left(\Lambda^t-\frac{\beta}{\alpha^2}C_t-q\frac{C_\phi}{\alpha}\right)
\mathbf{K}\left(1-\frac{y_m-y_n}{y_p-y_n}\right) +\right.
\\ \nn &&\left. (1-q^2)\Lambda^t \left(y_n \ \mathbf{K}\left(1-\frac{y_m-y_n}{y_p-y_n}\right)
+(y_p-y_n)\ \mathbf{E}\left(1-\frac{y_m-y_n}{y_p-y_n}\right)\right)\right],
\\ \label{Ss} &&S= \frac{2 T}{\sqrt{(1-q^2)\left[\left(\Lambda^\phi\right)^2
-\left(\Lambda^t\right)^2\right](y_p-y_n)}}
\\ \nn &&\left[\left(\frac{\beta}{\alpha^2}C_\phi+q\frac{C_t}{\alpha}+\Lambda^\phi q^2\right)
\mathbf{K}\left(1-\frac{y_m-y_n}{y_p-y_n}\right) +\right.
\\ \nn &&\left. (1-q^2)\Lambda^\phi \left(y_n \ \mathbf{K}\left(1-\frac{y_m-y_n}{y_p-y_n}\right)
+(y_p-y_n)\ \mathbf{E}\left(1-\frac{y_m-y_n}{y_p-y_n}\right)\right)\right.
\\ \nn &&\left. -\frac{q\frac{C_t}{\alpha}+q^2\Lambda^\phi}{1+y_p}\
\mathbf{\Pi}\left(\frac{y_p-y_m}{1+y_p},1-\frac{y_m-y_n}{y_p-y_n}\right)\right],\eea
where $\mathbf{K}$, $\mathbf{E}$ and $\mathbf{\Pi}$ are the complete elliptic integrals of first, second and third kind.

Now we turn to the $S^3$ subspace. By using
(\ref{SP}), (\ref{idc}) and (\ref{bfs}),
one can show that the scalar potential $U_\theta$ in (\ref{xi}) can be written as
\bea\label{Ux} &&U_\theta(\theta) = \frac{1}{2(\alpha^2-\beta^2)}
\left[\frac{\left(C_{\phi_2}-q\alpha\Lambda^{\phi_1}\ \chi\right)^2}{\chi}
+\frac{\left(C_{\phi_1}+q\alpha\Lambda^{\phi_2}\ \chi\right)^2}{1-\chi}\right.
\\ \nn &&\left. +\alpha^2 \left(\Lambda^{\phi_2}\right)^2 \chi+
\alpha^2\left(\Lambda^{\phi_1}\right)^2(1-\chi)\right],\eea
where we introduced the notation
\bea\label{chidef} \chi\equiv \cos^2\theta.\eea
Replacing (\ref{Ux}) in (\ref{xi}), one can see that it can be written in the form
\bea\label{xiS3} d\xi =\frac{\alpha^2-\beta^2}{2\alpha\sqrt{(1-q^2)
\left(\left(\Lambda^{\phi_1}\right)^2-\left(\Lambda^{\phi_2}\right)^2\right)}}
\frac{d\chi}{\sqrt{(\chi_p-\chi)(\chi-\chi_m)(\chi-\chi_n)}},\eea
where
\bea\nn 0\leq \chi_m<\chi<\chi_p\leq 1,\h \chi_n\leq 0,\eea
and
\bea\nn &&\chi_p+\chi_m+\chi_n =  \frac{1}{\alpha^2(1-q^2)
\left(\left(\Lambda^{\phi_1}\right)^2-\left(\Lambda^{\phi_2}\right)^2\right)}
\\ \nn &&
\left[-C_\theta(\alpha^2-\beta^2)
-\left(\alpha\Lambda^{\phi_2}\right)^2+
(2-q^2)\left(\alpha\Lambda^{\phi_1}\right)^2 \right.
\\ \nn &&\left.- 2q\alpha \left(C_{\phi_2}\Lambda^{\phi_1}
+C_{\phi_1}\Lambda^{\phi_2}\right)\right],\eea

\bea\nn &&\chi_p\chi_m+\chi_p\chi_n+\chi_m\chi_n=
 \frac{1}{\alpha^2(1-q^2)
\left(\left(\Lambda^{\phi_1}\right)^2-\left(\Lambda^{\phi_2}\right)^2\right)}
\\ \nn &&
\left[\left(\alpha\Lambda^{\phi_1}\right)^2+C_{\phi_1}^2
-C_{\phi_2}^2-C_\theta (\alpha^2-\beta^2)
- 2q\alpha C_{\phi_2}\Lambda^{\phi_1}\right],\eea
\bea\nn \chi_p\chi_m\chi_n=-\frac{\left(C_{\phi_2}\right)^2}{\alpha^2(1-q^2)
\left(\left(\Lambda^{\phi_1}\right)^2-\left(\Lambda^{\phi_2}\right)^2\right)}.\eea

Integrating (\ref{xiS3}), one finds the following solution for $\chi$
\bea\label{chisol} \chi(\xi)=(\chi_p-\chi_n)
\ \mathbf{DN}^2\left[\frac{\alpha\sqrt{(1-q^2)
\left(\left(\Lambda^{\phi_1}\right)^2-\left(\Lambda^{\phi_2}\right)^2\right)(\chi_p-\chi_n)}}
{\alpha^2-\beta^2}\ \xi,
\frac{\chi_p-\chi_m}{\chi_p-\chi_n}\right]+\chi_n.\eea

Now we are ready to find the ``orbit'' $r=r(x)$. Written in terms of $y$ and $\chi$,
it is given by
\bea\label{orbit} &&y= (y_p-y_n)\ \mathbf{DN}^2\left[\frac{\sqrt{
\left(\left(\Lambda^\phi\right)^2-\left(\Lambda^t\right)^2\right)(y_p-y_n)}}
{\sqrt{\left(\left(\Lambda^{\phi_1}\right)^2-\left(\Lambda^{\phi_2}\right)^2\right)(\chi_p-\chi_n)}}\right.
\\ \nn &&\left.\times \mathbf{F}\left(\arcsin\sqrt{\frac{\chi_p-\chi}{\chi_p-\chi_m}},
\frac{\chi_p-\chi_m}{\chi_p-\chi_n}\right),
\frac{y_p-y_m}{y_p-y_n}\right]+y_n
.\eea

Next, we compute $\tilde{X}^{\phi_1}(\xi)$ and $\tilde{X}^{\phi_2}(\xi)$.
Replacing the results in our ansatz,
we derive the following solutions for the isometric coordinates on $S^3$

\bea\label{thetatsol} &&\phi_1 = \Lambda^{\phi_1}\tau+
\frac{1}{\alpha\sqrt{(1-q^2)
\left(\left(\Lambda^{\phi_1}\right)^2-\left(\Lambda^{\phi_2}\right)^2\right)(\chi_p-\chi_n)}}
\\ \nn &&\left[\left(\beta\Lambda^{\phi_1}
-q\alpha\Lambda^{\phi_2}\right)\
\mathbf{F}\left(\arcsin\sqrt{\frac{\chi_p-\chi}{\chi_p-\chi_m}},\frac{\chi_p-\chi_m}{\chi_p-\chi_n}\right)\right.
\\ \nn &&\left. +\frac{C_{\phi_1}+q\alpha\Lambda^{\phi_2}}{1-\chi_p}
\ \mathbf{\Pi}\left(\arcsin\sqrt{\frac{\chi_p-\chi}{\chi_p-\chi_m}},-\frac{\chi_p-\chi_m}{1-\chi_p},
\frac{\chi_p-\chi_m}{\chi_p-\chi_n}\right)\right].\eea

\bea\label{thetasol} &&\phi_2 =\Lambda^{\phi_2}\tau+
\frac{1}{\alpha\sqrt{(1-q^2)
\left(\left(\Lambda^{\phi_1}\right)^2-\left(\Lambda^{\phi_2}\right)^2\right)(\chi_p-\chi_n)}}
\\ \nn &&\left[\left(\beta\Lambda^{\phi_2}
-q\alpha\Lambda^{\phi_1}\right)\
\mathbf{F}\left(\arcsin\sqrt{\frac{\chi_p-\chi}{\chi_p-\chi_m}},\frac{\chi_p-\chi_m}{\chi_p-\chi_n}\right)\right.
\\ \nn &&\left. +\frac{C_{\phi_1}}{\chi_p}
\ \mathbf{\Pi}\left(\arcsin\sqrt{\frac{\chi_p-\chi}{\chi_p-\chi_m}},1-\frac{\chi_m}{\chi_p},
\frac{\chi_p-\chi_m}{\chi_p-\chi_n}\right)\right],\eea

Based on (\ref{Qt}) and the solutions for the string coordinates on $S^3$ we found,
we can write down the explicit expressions for the conserved angular momenta $J_1$ and $J_2$
computed on the solutions. The result is

\bea\label{Jtsol} &&J_1 = \frac{2T}{\sqrt{(1-q^2)
\left(\left(\Lambda^{\phi_1}\right)^2-\left(\Lambda^{\phi_2}\right)^2\right)(\chi_p-\chi_n)}}
\\ \nn &&\left[\left(\frac{\beta}{\alpha^2}C_{\phi_1}+\Lambda^{\phi_1}
-q\frac{C_{\phi_2}}{\alpha}\right)
\mathbf{K}\left(1-\frac{\chi_m-\chi_n}{\chi_p-\chi_n}\right)\right.
\\ \nn &&\left.
-(1-q^2)\Lambda^{\phi_1}
\left(\chi_n\ \mathbf{K}\left(1-\frac{\chi_m-\chi_n}{\chi_p-\chi_n}\right)
+(\chi_p-\chi_n)\ \mathbf{E}\left(1-\frac{\chi_m-\chi_n}{\chi_p-\chi_n}\right)\right)\right].\eea

\bea\label{Jsol} &&J_2= \frac{2T}{\sqrt{(1-q^2)
\left(\left(\Lambda^{\phi_1}\right)^2-\left(\Lambda^{\phi_2}\right)^2\right)(\chi_p-\chi_n)}}
\\ \nn &&\left[\left(\frac{\beta}{\alpha^2}C_{\phi_2}-q\left(\frac{C_{{\phi_1}}}{\alpha}
+q\Lambda^{\phi_2}\right)\right)
\mathbf{K}\left(1-\frac{\chi_m-\chi_n}{\chi_p-\chi_n}\right)\right.
\\ \nn &&\left.
+(1-q^2)\Lambda^{\phi_2}
\left(\chi_n\ \mathbf{K}\left(1-\frac{\chi_m-\chi_n}{\chi_p-\chi_n}\right)
+(\chi_p-\chi_n)\ \mathbf{E}\left(1-\frac{\chi_m-\chi_n}{\chi_p-\chi_n}\right)\right)\right.
\\ \nn &&\left.
+\frac{q\left(\frac{C_{\phi_1}}{\alpha}+q\Lambda^{\phi_2}\right)}{1-\chi_p}
\ \mathbf{\Pi}\left(-\frac{\chi_p-\chi_m}{1-\chi_p},
1-\frac{\chi_m-\chi_n}{\chi_p-\chi_n}\right)\right],\eea

Now, let us go to the $T^4$ subspace.
Since in terms of $\varphi^i$ coordinates the metric is flat and there is no $B$-field,
the solutions for the string coordinates are simple and given by
\bea\label{fi} \varphi^i(\tau,\sigma)= \Lambda^i\tau
+\frac{1}{\alpha^2-\beta^2} \left(C_i
+\beta\Lambda^i\right)\xi.\eea
The conserved charges (\ref{Ji}) can be computed to be
\bea\label{Jisol} J_i^T=\frac{2\pi\alpha T}{\alpha^2-\beta^2}
\left(\frac{\beta}{\alpha}C_i+\alpha\Lambda^i\right).\eea

If we impose the periodicity conditions
\bea\nn  \varphi^i(\tau,\sigma)= \varphi^i(\tau,\sigma+2L)+2\pi n_i,\h n_i\in \mathbb{Z}_i,\eea
the integration constants
$C_i$ are fixed in terms of the embedding parameters. Namely,
\bea\label{Ci} C_i= \frac{\pi n_i}{L\alpha}(\alpha^2-\beta^2)
-\beta\Lambda^i.\eea
Replacing (\ref{Ci}) into (\ref{fi}) and (\ref{Jisol}), one finally finds
\bea\label{T4solJi} &&\varphi^i =\left(\Lambda^i+\frac{\beta}{\alpha}
\frac{\pi n_i}{L}\right)\tau
+\frac{\pi n_i}{L}\sigma,
\\ \nn &&J_i^T =2\pi T \left(\Lambda^i+\frac{\beta}{\alpha}\frac{\pi n_i}{L}\right).\eea

Let us finally point out that the Virasoro constraints impose the following two conditions
on the embedding parameters and integrations constants in the solutions found
\bea\label{Vc} &&C_r+C_\theta=0,
\\ \nn &&\Lambda^t C_t+\Lambda^\phi C_\phi+\Lambda^{\phi_1} C_{\phi_1}+\Lambda^{\phi_2} C_{\phi_2}
-\Lambda^i\left(\beta\Lambda^i-(\alpha^2-\beta^2)\frac{\pi n_i}{\alpha L}\right)=0.\eea

\end{appendix}


\begin{thebibliography}{99}

\bibitem{AdS/CFT} J. M. Maldacena, ``The large N limit of superconformal
field theories and supergravity'', Adv. Theor. Math. Phys. {\bf 2},
231 (1998) [{arXiv:hep-th/9711200}];\\S. S. Gubser, I. R. Klebanov
and A. M. Polyakov, ``Gauge theory correlators from non-critical
string theory'',
Phys. Lett. {\bf B428}, 105 (1998) [{arXiv:hep-th/9802109}];\\
E. Witten, ``Anti-de Sitter space and holography'', Adv. Theor.
Math. Phys. {\bf 2}, 253 (1998) [{arXiv:hep-th/9802150}].

\bibitem{RO} Niklas Beisert, Changrim Ahn, Luis F. Alday, Zoltan Bajnok, James M. Drummond,
Lisa Freyhult, Nikolay Gromov, Romuald A. Janik, Vladimir Kazakov,
Thomas Klose, Gregory P. Korchemsky, Charlotte Kristjansen, Marc
Magro, Tristan McLoughlin, Joseph A. Minahan, Rafael I. Nepomechie,
Adam Rej, Radu Roiban, Sakura Schafer-Nameki, Christoph Sieg,
Matthias Staudacher, Alessandro Torrielli, Arkady A. Tseytlin, Pedro
Vieira, Dmytro Volin, Konstantinos Zoubos, ``Review of AdS/CFT
Integrability: An Overview'', Lett. Math. Phys. {\bf 99} 3 (2012)
[arXiv:hep-th/1012.3982v5].

\bibitem{BSZ0912} A. Babichenko, B. Stefanski, K. Zarembo,
``Integrability and the AdS(3)/CFT(2) correspondence'',
 JHEP 1003 (2010) 058, [arXiv:hep-th/0912.1723v3].

\bibitem{SS1106} O. Ohlsson Sax, B. Stefanski Jr,
``Integrability, spin-chains and the AdS3/CFT2 correspondence'',
JHEP 1108 (2011) 029 [arXiv:hep-th/1106.2558v2].

\bibitem{KW1106} Ingo Kirsch, Tim Wirtz,
``Worldsheet operator product expansions and p-point functions in
AdS3/CFT2'', JHEP 1110 (2011) 049 [arXiv:hep-th/arXiv:1106.5876v2].

\bibitem{RSW1204} Nitin Rughoonauth, Per Sundin, Linus Wulff,
``Near BMN dynamics of the AdS(3) x S(3) x S(3) x S(1)
superstring'', JHEP 1207 (2012) 159 [arXiv:hep-th/1204.4742v3].

\bibitem{CZ1209} Alessandra Cagnazzo, Konstantin Zarembo,
``B-field in AdS(3)/CFT(2) Correspondence and Integrability'',
 JHEP 1211 (2012) 133,
[arXiv:hep-th/1209.4049v2].

\bibitem{SST1211} Olof Ohlsson Sax, Bogdan Stefanski jr, Alessandro
Torrielli, ``On the massless modes of the AdS3/CFT2 integrable
systems'', JHEP 1303 (2013) 109,
[arXiv:hep-th/1211.1952v2].

\bibitem{CB1211} Changrim Ahn, Diego Bombardelli,
``Exact S-matrices for $AdS_3/CFT_2$'', [arXiv:hep-th/1211.4512].

\bibitem{BSS1211} Riccardo Borsato, Olof Ohlsson Sax, Alessandro
Sfondrini, ``A dynamic $su(1|1)^2$ S-matrix for AdS3/CFT2'', JHEP
1304 (2013) 113 [arXiv:hep-th/1211.5119v3].

\bibitem{BLMT1211} M. Beccaria, F. Levkovich-Maslyuk, G. Macorini, A. A.
Tseytlin, ``Quantum corrections to spinning superstrings in $AdS_3
\times S^3 \times M^4$: determining the dressing phase'', JHEP 1304
(2013) 006, [arXiv:hep-th/1211.6090v3].

\bibitem{BSS1212} Riccardo Borsato, Olof Ohlsson Sax, Alessandro
Sfondrini,``All-loop Bethe ansatz equations for AdS3/CFT2'', JHEP
1304 (2013) 116, [arXiv:hep-th/1212.0505v3].

\bibitem{BM1212} Matteo Beccaria, Guido Macorini,
``Quantum corrections to short folded superstring in $AdS_3 \times
S^3\times M^4$'', JHEP 1303 (2013) 040, [arXiv:hep-th/1212.5672v2].

\bibitem{SW1302} Per Sundin, Linus Wulff, ``Worldsheet scattering in
AdS(3)/CFT(2)'', JHEP 1307 (2013) 007 [arXiv:hep-th/1302.5349v2].

\bibitem{HT1303} B. Hoare, A. A. Tseytlin,
``On string theory on $AdS_3 \times S^3 \times T^4$ with mixed
3-form flux: tree-level S-matrix'', Nucl.Phys. B873 (2013) 682-727,
[arXiv:hep-th/1303.1037v4].

\bibitem{BSSST1303} Riccardo Borsato, Olof Ohlsson Sax, Alessandro Sfondrini, Bogdan Stefanski, Alessandro
Torrielli, ``The all-loop integrable spin-chain for strings on
$AdS_3 \times S^3 \times T^4$: the massive sector'', JHEP 1308
(2013) 043, [arXiv:hep-th/1303.5995v2].

\bibitem{HT1304} B. Hoare, A. A. Tseytlin,
``Massive S-matrix of $AdS_3 \times S^3 \times T^4$ superstring
theory with mixed 3-form flux'', Nucl.Phys. B873 (2013) 395-418 ,
[arXiv:hep-th/1304.4099v3].

\bibitem{EMR1304} Oluf Tang Engelund, Ryan W. McKeown, Radu Roiban,
``Generalized unitarity and the worldsheet S matrix in $AdS_n \times
S^n \times M^(10-2n)$'',  JHEP 1308 (2013) 023
,[arXiv:hep-th/1304.4281v1].

\bibitem{BSSST1306} Riccardo Borsato, Olof Ohlsson Sax, Alessandro Sfondrini,
Bogdan Stefanski, Alessandro Torrielli, ``Dressing phases of AdS3/CFT2'', Phys.Rev. D88 (2013)
066004, [arXiv:hep-th/1306.2512v2].

\bibitem{A1306} Michael C. Abbott,
``The $AdS3 \times S3 \times S3 \times S1$ Hernandez-Lopez Phases: a
Semiclassical Derivation'', J. Phys. A46 (2013) 445401 [arXiv:hep-th/1306.5106v2].

\bibitem{SW1306} Per Sundin, Linus Wulff,
``The low energy limit of the $AdS(3)\times S(3)\times M(4)$
spinning string'', JHEP 1310 (2013) 111 [arXiv:hep-th/1306.6918v1].

\bibitem{HST1311} B. Hoare, A. Stepanchuk, A.A. Tseytlin,
``Giant magnon solution and dispersion relation in string theory in
$AdS_3 \times S^3 \times T^4$ with mixed flux'', [arXiv:hep-th/1311.1794].

\bibitem{LS1312} Thomas Lloyd, Bogdan Stefański jr, ``AdS 3 /CFT 2,
finite-gap equations and massless modes'', arXiv:1312.3268 [hep-th]

\bibitem{HM06} Diego M. Hofman, Juan Maldacena, ``Giant Magnons'', J.Phys.A39:13095-13118,2006
[arXiv:hep-th/0604135].

\bibitem{CDO06} Heng-Yu Chen, Nick Dorey, Keisuke Okamura, ``Dyonic Giant
Magnons'', JHEP 0609 (2006) 024, [arXiv:hep-th/0605155].

\bibitem{AFZ06} Gleb Arutyunov, Sergey Frolov, Marija Zamaklar,
``Finite-size Effects from Giant Magnons'',  Nucl.Phys. B778 (2007)
1-35, [arXiv:hep-th/0606126].

\bibitem{HS08} Yasuyuki Hatsuda, Ryo Suzuki,
``Finite-Size Effects for Dyonic Giant Magnons'', Nucl.Phys. B800
(2008) 349-383, [arXiv:hep-th/0801.0747].










\end{thebibliography}
\end{document}